\documentclass{jetpl}
\twocolumn

\lat

\DeclareMathOperator{\Tr}{Tr}
\def\omir{\omega_\mathrm{ir}}
\def\omc{\omega_\mathrm{c}}

\title{Dephasing of Josephson qubits close to optimal points}

\rtitle{Dephasing of Josephson qubits close to an optimal point\ldots}

\sodtitle{Dephasing of Josephson qubits in the vicinity of optimal points}

\author{S.\,V.\,Syzranov$^{+,*}$
\/\thanks{e-mail: syzranov,\,makhlin@itp.ac.ru}
and Yu.~Makhlin$^+$
}

\rauthor{S.\,V.\,Syzranov and Yu.~Makhlin}

\sodauthor{Syzranov, Makhlin}

\address{
$^+$Landau Institute for Theoretical Physics, Kosygin st. 2,
119334 Moscow, Russia\\~\\
$^*$ 
Moscow Institute of Physics and Technology, 
 Institutskiy prosp. 9, Dolgoprudniy, 141700 Moscow Region, Russia
}

\dates
{
21 December 2005
}
{
*
}

\abstract{Decoherence of Josephson qubits can be substantially reduced by tuning
their parameters to optimal operation points, with only quadratic coupling to fluctuations.
We analyze dephasing due to $1/f$ noise for a two-level system, detuned from
an optimal point, i.e., the crossover to the linear-coupling regime,
both for free induction decay and for spin-echo experiments.
Influence of several noise sources is also discussed.}


\begin{document}

\maketitle

Superconducting nanocircuits are promising candidates for
implementation of quantum-coherent two-level systems, building blocks of
prospective quantum information processing
devices~\cite{Esteve:review,Devoret:review}.
At present, coherence of these circuits is limited mostly by low-frequency
noise, often with $1/f$ power spectrum, such as background-charge
fluctuations, variations of magnetic flux in superconducting loops, and of the
critical current of Josephson junctions~\cite{Ithier:quantronium2,
 Martinis:phase2, Bertet:flux}.

An efficient method of improving the coherence properties of qubits, the ``optimal
point strategy'', was suggested in~\cite{Vion:quantronium}: the bias
of a charge-phase qubit was tuned to an operation point where the linear coupling to the noise
sources vanishes. As a result, the coherence time was extended by 2--3
orders of magnitude compared to earlier experiments. This approach was 
later applied to reduce the influence of bias-current fluctuations on a
flux qubit~\cite{Bertet:flux}.
This method may be used in combination with the spin-echo-type
techniques~\cite{Ithier:quantronium2,Bertet:flux,Nakamura:charge}, inherited from NMR.

The long coherence time in these devices allows to generate long-lived
quantum-coherent oscillations and to study their decay laws in detail.
Thus solid-state qubits may be used as a unique tool to gather information
about the properties of the noise~\cite{Aguado:detector,Schoelkopf:detector}.
Such studies~\cite{Ithier:quantronium2,Bertet:flux,Astafiev:strange} allow one to
investigate the origin and the microscopic mechanism of the noise and to
design devices with even better coherence, necessary for
large-scale quantum circuits.

Analysis of decoherence at and close to an optimal point is important for
superconducting quantum bits. Comparison of dephasing at an optimal point and
away provides additional information on the nature and statistical properties
of the fluctuations (cf.~Ref.~\cite{Ithier:quantronium2}). Furthermore,
it is relevant since a two-level system may be tuned away from degeneracy for
quantum manipulations or readout. In other words, here we analyze the
crossover between the optimal bias conditions and the linear-coupling regime
far away from degeneracy.

Recent experiments suggest that the low-frequency noise in Josephson circuits
is produced by collections of bistable fluctuating systems, a well-known model
of the flicker noise~\cite{Dutta:review}. For large collections with a
sufficiently regular distribution of parameters, one expects the noise to obey
Gaussian statistics due to the central limit theorem. For small number of
fluctuators or singular distributions, non-Gaussian effects may strongly
influence decoherence~\cite{Paladino:fluctuators,GalAltShan}.

Dephasing of qubits by long-correlated Gaussian noise at optimal points was
studied earlier for $1/f$ noise spectrum in the cases of free-induction
decay~\cite{Makhlin:quadrdeph} and echo~\cite{Schriefl}
experiments (cf. also Ref.~\cite{Rabenstein:lorentz}). Here we analyze decoherence
in a system subject to Gaussian $1/f$ noise, from one or several independent
sources, in the vicinity of an optimal point, such that both quadratic and
linear coupling terms are relevant. 

We express the dephasing law of a system near an optimal point in terms of the
decay law precisely at the optimal point. We show that the ratio of these two
dephasing laws is dominated by the low-frequency noise. This allows us to find
the relevant laws for free-induction and spin-echo decay. We also generalize
the analysis to the case of several noise sources.

Below we assume that the infrared cutoff frequency of $1/f$ noise is the
lowest frequency scale in the problem, and in particular, is much lower than $1/t$
for relevant times. In our analysis we exploit this fact shifting, under
proper conditions, the spectral weights of the fluctuations between zero
frequency and $\omir$.

We consider a two-level system (spin-1/2) with the Hamiltonian
\begin{equation}\label{Ham}
{\cal H} = -\frac{1}{2} \left[ \varepsilon_0+\vartriangle \varepsilon(X(t))\right]
\hat\sigma_z
+ {\cal H}_\mathrm{bath} \,,
\end{equation}
which describes coupling to a noise source via a fluctuating quantity $X(t)$.
In turn, its dynamics is governed by the Hamiltonian of the noise source, ${\cal
  H}_\mathrm{bath}$.
Physically, $X(t)$ may represent, e.g., gate-voltage or magnetic-flux fluctuations in
a Josephson qubit. We assume the $1/f$ spectrum of its fluctuations in the
relevant frequency range.

Eq.~(\ref{Ham}) presents the Hamiltonian in the eigenbasis of the
non-perturbed ($\Delta\varepsilon=0$) part. Only the fluctuations of the
level splitting, i.e., the diagonal terms, are accounted for. Note that as far
as the influence of the low-frequency noise ($\omega\ll\varepsilon_0$) on the
qubit is concerned, the general case reduces to Eq.~(\ref{Ham}) in the
adiabatic approximation~\cite{Makhlin:transverse}.

To characterize decoherence, we analyze the decay of the off-diagonal entry of
the qubit's density matrix in the eigenbasis. Averaging over the noise
realizations, we find
\begin{eqnarray}
	\langle\widehat{\sigma}_-(t)\rangle\equiv
	 \Tr\left(\widehat{\sigma}_-\widehat{\rho}(t) \right)=
	\langle\widehat{\sigma}_-(0)\rangle
	e^{it\varepsilon_0}P(t), \nonumber \\
	P(t)=\langle \mathrm{T}\,
	 e^{i\int_{0}^{t}g(t)\vartriangle\varepsilon(X(t))dt} \rangle,
	\label{def}
\end{eqnarray}
where $g(t)\equiv 1$ for the free induction decay. More generally, the
function $g(t)$ accounts for modulations of the qubit-noise coupling.
For example, in echo-type experiments $|g(t)|=1$ and the sign of $g(t)$
is reversed every time when a $\pi$-pulse is applied.

In the vicinity of an optimal point, keeping the leading-order terms in the
expansion of $\Delta\varepsilon$ in $X$, we find
\begin{eqnarray}
	\vartriangle\varepsilon\left(X(t)\right)
	=\lambda\left(X(t)+\frac{D}{2}\right)^2,
\end{eqnarray}
where we have combined the linear and quadratic terms to form a full square,
by transferring a constant from $\varepsilon_0$ for convenience.
Here the constant $D$ characterizes the offset from the optimal point and
vanishes at the degeneracy.

{\it Dependence on the shift from the optimal point.\/}
One can represent the averaging in Eq.~(\ref{def}) via a Gaussian functional
integral~\cite{Schriefl} over $X(t)$,
of the type $\int e^{-Y^TAY+D(\xi^TY+Y^T\xi)+cD^2}dY$. This implies a
relation between the
decay law of coherence in the vicinity of the optimal point, $P(t)$, and the
dephasing law $P_0(t)$ at the optimal point (i.e., for $D=0$):
\begin{eqnarray}
	P(t)=P_0(t)e^{-\lambda D^2f(t)}
	\label{fdef}
\end{eqnarray}
with a $D$-independent function $f(t)$. Naturally, this function
depends on $g(t)$ and on the noise power $S_X(\omega)$.
Eq.~(\ref{fdef}) also follows from the diagrammatic
  analysis~\cite{Makhlin:quadrdeph,Makhlin:transverse}: circular diagrams, considered in
  Ref.~\cite{Makhlin:quadrdeph}, contribute to $P_0(t)$, and (the logarithm of) the
  second term is a combination of linear diagrams, with two linear-coupling
  vertices at the ends (Fig.~1), thus leading to the $D^2$ dependence.
Below we find $f(t)$ in various situations. Note that in the far-detuned limit
$D\to\infty$ only the short-time behavior of $f(t)$ is relevant for the
dephasing law.

\begin{figure}
\centerline{\includegraphics[width=6.0cm]{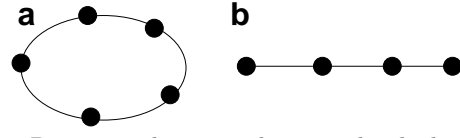}}
\caption[]{\label{F:diag}FIG.\ref{F:diag}.
Diagrams that contribute to the dephasing at an optimal point (a) and close to
it (b)}
\end{figure}

{\it Reduction to the dephasing law at the optimal point.}
Let us formally average the last equation over a Gaussian-distributed $D$ with
dispersion $\sigma^2$. Clearly, the average of the dephasing law (\ref{fdef})
over this distribution coincides with the dephasing induced by fluctuations
with the modified noise power, $\tilde S(\omega)=S(\omega) +
{\textstyle\frac{\pi}{2}} \sigma^2 \delta(\omega)$.

Physically, at the time scale $t$ the low-frequency part of $X(t)$ (i.e.,
the contribution of the frequencies $\omega\ll 1/t$) behaves as a static
($\omega=0$) random Gaussian quantity, and the dephasing law depends only
on its dispersion, i.e., the low-frequency weight $\int d\omega
S(\omega)$. Technically, this follows, for instance, from the diagrammatic
analysis, similar to that in Ref.~\cite{Makhlin:quadrdeph}.
In particular, the addition of a new component at zero frequency is equivalent
to the renormalization $\omir\to\omir'$:
\begin{equation}\label{omir-renorm}
\lambda\frac{\sigma^2}{4}+\frac{\Gamma}{\pi}|\ln(t\omir)|=
\frac{\Gamma}{\pi}|\ln(t\omir')| \,,
\end{equation}
where $\Gamma=\lambda X_f^2$, and $X_f$ characterizes the strength of the $1/f$ noise,
\begin{equation}
S(\omega)=\frac{X_f^2}{\mid\omega\mid}.
\end{equation}

Thus after averaging (\ref{fdef}) over $D$, one obtains 
\begin{equation}
P_0(t,\omega_{ir}e^{-\frac{\pi\lambda}{4\Gamma}\sigma^2})=
\frac{P_0(t,\omir)}{\left(1+2\lambda\sigma^2f(t)\right)^{1/2}}
\,.
\label{rel}
\end{equation}
Comparing the subleading terms in the expansion in $\sigma^2$ at $\sigma\to
0$, we find~\footnote
{
For an arbitrary (non-$1/f$) spectrum with a sharp low-frequency cutoff at a
certain frequency $\omir$ one finds $f(t)=\frac{\pi}{4\lambda
  S(\omir)}\frac{\partial\ln P_0}{\partial\omir}$.
}

\begin{equation}
f(t)=\frac{\pi}{4\Gamma}\frac{\partial\ln P_0(t)}{\partial\ln\omir}.
\label{main}
\end{equation}

Taking into account the dimensions of time $t$, noise power $\Gamma$,
infrared $\omir$ and ultraviolet $\omc$ cutoff frequencies,
 one can rewrite Eq.~(\ref{rel}) as
\begin{eqnarray}
	f(t)=
	\frac{\pi}{4\Gamma}\left(\frac{\partial\ln P_0(t)}{\partial\ln t}-
	\frac{\partial\ln P_0(t)}{\partial\ln \Gamma}
	-\frac{\partial\ln P_0(t)}{\partial\ln \omc}\right).
\end{eqnarray}

Eqs.~(\ref{fdef}) and (\ref{main}) allow one to find
the dephasing close to an optimal point in terms of the dephasing law
precisely at this optimal point.

{\it Free induction decay, high ultraviolet cutoff.}
Short-time asymptotics of $P_0(t)$ were found earlier~\cite{Makhlin:quadrdeph,
  Schriefl}. For example, in the case of free induction decay and
high ultraviolet cutoff, $\omc t\gg1$, one immediately finds from
Eqs.~(\ref{fdef}), (\ref{main}) and the results of
Ref.~\cite{Makhlin:quadrdeph} that
\begin{eqnarray}
	P(t)&=&\left( 1-\left(\frac{2}{\pi}i\Gamma t
	\ln\frac{1}{\omir t}\right)\right)^{-1/2}\cdot e^{-\lambda D^2f(t)},
	\label{Pshort} \\[3mm]
	f(t)&=&\frac{it}{4}\frac{1}{\frac{2}{\pi}i\Gamma t\ln\frac{1}{\omir t}-1}.
	\label{freemain}
\end{eqnarray}
These results are based on the short-time behavior of $P_0(t)$ at $\Gamma
t\ll1$ (and $|\ln(\omir t)|\gg1$)~\cite{Makhlin:quadrdeph}.
They are dominated by the contribution of the low-frequency fluctuations
($|\omega|\ll 1/t$). We show below that this contribution dominates at longer
times as well; thus Eq.~(\ref{freemain}) (but not Eq.~(\ref{Pshort})) applies at longer times too.

Indeed, we have seen that $f(t)$ can be found from the sensitivity of $P_0(t)$
to $\omir$. Clearly, the contribution of high frequencies $\omega\gtrsim 1/t$
is insensitive to the infrared cutoff. Before providing an accurate evaluation
of $f(t)$ let us remark that its low-frequency contribution can be estimated
by treating the low-frequency part as static noise with the same dispersion,
$\sigma^2=\frac{X_f^2}{\pi} \ln\frac{1}{\omir t}$. Then, averaging the phase factor
$\exp(i\lambda(X+D/2)^2)$ over the Gaussian distribution $\propto
\exp(-X^2/2\sigma^2)$, we find immediately the results (\ref{Pshort}) and
(\ref{freemain}).

To calculate $f(t)$ one has to evaluate the diagrams in Fig.~1b. In
contrast to the circular diagrams in Fig.~1a, the high-frequency
contributions in Fig.~2b are suppressed since the incoming
frequencies at the ends are zero, and at each vertex the typical frequency
change is of order $1/t$. Accurate evaluation demonstrates that this indeed
suppresses all contributions apart from (\ref{freemain}), dominated by very
low frequencies $\omega\ll 1/t$.

We remark that in the far-detuned limit $D^2\to\infty$ the total decay
$P_0(t)\exp[-\lambda D^2 f(t)]$ is dominated by the second factor and by the
short-time asymptotics of $f(t)$. In this limit one recovers, as expected, the
decay laws for a linearly coupled reservoir.

{\it Free induction decay, low ultraviolet cutoff.}
For $\omc t\lesssim1$, $X(t)$ can be treated as static, and one easily gets 
\begin{eqnarray}
	P_0(t)=\left( 1-\frac{2}{\pi}it\Gamma 
	\ln\frac{\omc}{\omir}\right)^{-1/2}, \\
	f(t)=\frac{it/4}{\frac{2}{\pi}it\Gamma\ln\frac{\omc}{\omir}-1}\,.
\end{eqnarray}

{\it Spin echo decay near optimal points.}
For the case of free induction decay, the short-time asymptotics of $P_0(t)$
gave via Eq.~(\ref{main}) an expression for $f(t)$ applicable at all relevant
times. Similarly, one can find $f(t)$ for a system subject to
spin-echo pulses, using Eq.~(\ref{main}) and the dephasing laws at the
optimal point in the corresponding cases~\cite{Schriefl} (for one or several
$\pi$-pulses and depending on the value and shape of the ultraviolet cutoff).

In particular, for $N-1$ spin-echo pulses ($N\geq2$) and a sufficiently high ultraviolet
cutoff, $\omc\gg N/t$, one finds
\begin{eqnarray}
	f(t)=\frac{\pi}{8\Gamma}
	\frac{\frac{C_{N}}{N}\left(\frac{2}{\pi}\Gamma t\right)^2}
	{1+\frac{C_{N}}{N}\left(\frac{2}{\pi}\Gamma t\right)^2\ln\frac{1}{t\omega_{ir}}}\,,
\end{eqnarray}
where $C_{N}$ is a constant of order 1 \cite{Schriefl}.

However, the derivation of Eq.~(\ref{main}) for spin-echo decay requires
stricter conditions: it implies  that the dephasing doesn't change
significantly if the offset $D$ from the
optimal point is substituted by a slow oscillating offset with frequency
$\omir\ll1/t$. As a result of such a substitution the term $\lambda\int
g(t)D^2dt\equiv0$ is substituted by a non-zero quantity. This does not change
the result as long as this quantity is much less than $\lambda D^2f(t)$. An
estimate of the inaccuracy of Eq.~(\ref{main}) leads to
the constraint $(t\omir)t\ll Nf(t)$ on its applicability to an echo-type
experiment considered, which for relevant experimental parameters
($\omir\ll\Gamma$) reduces to 
$(t\omir)t\ll N/|\Gamma \ln(t\omir)|$.
 For time scales of interest in spin-echo experiments
this constraint is satisfied.

{\it Several fluctuating parameters.}
Typically, the behavior of Josephson qubits is controlled by several
parameters, exhibiting low-frequency noise.
The results derived above can be generalized to the situation with many 
 sources of $1/f$ noise in the vicinity of their optimal points.

In the simplest case, various noise sources contribute independently:
\begin{eqnarray}
	\vartriangle\varepsilon=\sum_i \lambda_iX_i^2+\sum_i \lambda_iX_i D_i.
\end{eqnarray}
Then the dephasing is a product of the partial contributions,
$P(t)=\prod_i P_i(t)$, regardless of the noise spectra. 

However, in general the quadratic part of the perturbation near the optimal
point contains cross terms, $X_iX_j$. For instance, this is the case for the
flux qubit in Ref.~\cite{Mooij:twoflux} and two fluxes as control parameters.
If all noise sources have similar spectra (such that
$S_i(\omega)=\mathrm{const}\cdot S_j(\omega)$), one can reduce the problem to that
for independent sources by simultaneous diagonalization of the quadratic
perturbation and the correlation matrix $\langle X_i(t)X_j(t')\rangle$
(cf.~Ref.~\cite{Schriefl}). For instance, this remark applies to several
sources of $1/f$ noise with the same infrared cutoff (and, e.g., sufficiently high
ultraviolet cutoffs, $\omc^i t\gtrsim1$, which do not influence the dephasing law
strongly).

In general, however, different sources of $1/f$ noise may have different
infrared cutoffs. Still, this case can be reduced to the problem of a single
noise source. Indeed, to treat this case, one can employ the approach used above
in the derivation of~(\ref{omir-renorm}): one adjusts all the infrared cutoff
frequencies, $\omir^i$, to the same value (still, $\ll 1/t$) by shifting the
rest of the low-frequency spectral weight to zero frequency. As a result, one
arrives at the situation with several noise sources with the same spectra (up
to an overall factor) and a static random field. At this point, one can
average first over the dynamical fluctuations by re-diagonalization as above,
and then over the static noise. Notice, that fluctuations with a low
ultraviolet cutoff, $\ll 1/t$, may be treated as low-frequency fluctuations in
the same spirit.

In particular, the dephasing for a linear combination
$X(t)=\sum\alpha_iX_i(t)$ of $1/f$ noise sources with high ultraviolet
cutoffs can be obtained following the procedure described.
 One finds that the dephasing in this case coincides
with the dephasing due to a single source of $1/f$ noise 
with the power $\Gamma=\sum|\alpha_i|^2\Gamma_i$ and the infrared cutoff
\begin{equation}
\omir=\Pi_{i}\left(\omir^i\right)^{\alpha_i^2\Gamma_i/\Gamma}
\,.
\end{equation}

In conclusion, we considered the dephasing of a two-level system
close to an optimal operation point, in the crossover regime between the
quadratic and the linear coupling to the noise source. 
We found the dependence on the shift from the optimal point, by relating the
dephasing law to that at the optimal point with only quadratic coupling.
In the case of $1/f$ noise, we present an explicit expression for the ratio of
two dephasing laws. Further, we analyzed the influence of several noise
sources and demonstrated, how the problem can be reduced to the case of a
single source.

We thank A.~Shnirman for useful discussions. This work was supported by the
Dynasty foundation and the grant MD-2177.2005.2.

\end{document}